\documentstyle[twoside,fleqn,espcrc2]{article}
\include{figures}
\title{{\bf The Strange-Quark Mass From QCD Sum Rules: An Update}} 
\author{
{\bf C. A. Dominguez}\\
\vspace{0.5cm}
Institut f\"{u}r Physik der Ludwig Maximilian Universit\"{a}t\\
M\"{u}nchen, Germany, and\\
Institute of Theoretical Physics and Astrophysics\\
University of Cape Town, South Africa
}
\begin{document}
\begin{abstract}
\noindent
Three different ways of determining the strange quark mass using QCD
sum rules are reviewed. First, from a QCD sum rule determination of
the up and down quark masses, together with the current algebra ratio
$ m_{s}/(m_{u}+m_{d})$. Second and third, from QCD sum rules involving 
the strangeness changing vector and axial-vector current divergences, 
respectively. Present results are encompassed in the value: 
$m_{s} \; \mbox{(1 GeV)} = 150 \pm \mbox{70 MeV}$. It is argued that until
{\it direct},
and precision measurements of the relevant spectral functions become
available, the above error is not likely to be reduced.
\end{abstract}
\maketitle
\setlength{\baselineskip}{1\baselineskip}
It is probably unnecessary, in such a workshop, to emphasize the importance
of knowing the value of the strange quark mass with reasonable accuracy.
Most of kaon-physics, plus CP violation, depend strongly on $m_{s}$.
Also, sophisticated modern multiloop calculations in weak hadronic physics 
will not achieve their full potential of
improving theoretical precision, unless the uncertainty on the value of
$m_{s}$ is brought under control. A couple of years ago, it seemed that
this was the case, as progress was made in understanding and removing
quark mass singularities from light-quark correlators 
\cite{CH1}-\cite{J1}. This allowed for a better control of the theoretical
(QCD) side
of the sum rules in e.g. the $\Delta S = 1$ scalar channel, where
experimental data on the $K-\pi$ phase shifts allows, in principle, to
{\it reconstruct} the hadronic spectral function. The word {\it reconstruct}
is crucial, since it is no substitute for a direct and precision measurement 
of the spectral function, as I shall argue here. In fact, a recent
re-analysis \cite{IT1} of these QCD sum rules, using the same  
$K-\pi$ phase shift primary data, but a different reconstruction
of the spectral function, has made this evident.\\  

I start with the determination of $m_{s}$ from the current algebra ratio
\cite{GL} 
\begin{equation}
\frac{m_{s}}{m_{u}+m_{d}} = 12.6 \pm 0.5 \; ,
\end{equation}
together with a QCD sum rule determination of ($m_{u}+m_{d}$). The latter
was discussed some years ago \cite{CAD1} in the framework of QCD Finite
Energy Sum Rules (FESR) in the pseudoscalar channel,
at the two-loop level in perturbative QCD,
and including non-perturbative condensates up to dimension-six, with
the result
\begin{equation}
(m_{u}+m_{d}) \; \mbox{(1 GeV)} = 15.5 \pm \mbox{2.0 MeV} \; .
\end{equation}
This result, together with Eq.(1), implies
\begin{equation}
m_{s} \; \mbox{(1 GeV)} = 195 \pm \mbox{28 MeV} .
\end{equation}
A recent re-analysis \cite{BIJ} of the same QCD sum rules, 
but including the next  (3-loop) order in perturbation theory, quotes
\begin{equation}
(m_{u}+m_{d}) \; \mbox{(1 GeV)} = 12.0 \pm \mbox{2.5 MeV} \; ,
\end{equation}
which, using Eq.(1), leads to
\begin{equation}
m_{s} \; \mbox{(1 GeV)} = 151 \pm \mbox{32 MeV} \; .
\end{equation}
The problem here is that essentially the same raw data for resonance masses
and widths, plus the same threshold normalization from chiral
perturbation theory, has been used in \cite{CAD1} and \cite{BIJ}. The
main difference is in the reconstruction of the spectral function from
the resonance data; with a more ornamental functional form  being adopted
in \cite{BIJ}. In fact, the difference between the results of both analyses
cannot be accounted for by the inclusion (or not) of the 3-loop perturbative
QCD contribution, which amounts to a reduction of
the 2-loop result, Eq.(2),
of only a few percent. The difference lies in the functional form of the
hadronic parametrization of the same set of raw resonance data. This
exposes a type of {\it systematic} uncertainty of the QCD sum rule
method, and will cease to be an uncertainty only after the pseudoscalar
hadronic spectral function is measured {\it directly}
and accurately (e.g. from tau-lepton decays). In the meantime, both
results for $m_{s}$, Eqs. (3) and (5), are equally acceptable, and
taken together provide a measure of underlying systematic uncertainties.\\

Next, I turn to the second alternative, which is based on the correlator
of the strangeness-changing vector current divergences
\begin{eqnarray}
\psi\;(q^2) = i\;\int\;d^4\;x\;e^{iqx} \nonumber\\
<0|T(\partial^{\mu}\;V_{\mu}(x)\;\partial^{\nu}
		 V_{\nu}^{\dagger}(0))|0> \; ,
\end{eqnarray}
where $V_{\mu}(x) = :\bar{s}(x)\gamma_{\mu} u(x):$, and
$\partial^{\mu}V_{\mu}(x) = m_{s} \; :\bar{s}(x) i u(x):$,
and the up and down quark masses have been neglected.
To summarize the QCD sum rule technique \cite{SVZ} for determining
$m_{s}$, one first makes use 
of Wilson's Operator
Product Expansion (OPE) of current correlators at short distances,
modified so as to include non-perturbative effects
(parametrized by quark and gluon vacuum condensates).
This provides a QCD, or theoretical expression
of the two-point function which involves the quark mass to be determined,
the QCD scale $\Lambda_{QCD}$, and the vacuum condensates (whose values
are determined independently).
Next, invoking analyticity and QCD-hadron
duality, the QCD correlator can be equated to a weighted integral of the
hadronic spectral function. Depending on the choice of weight, one
obtains e.g. FESR, Laplace and Hilbert
transform sum rules, Gaussian sum rules, etc..
The upper limit of the hadronic integral, the so called continuum (or
asymptotic freedom) threshold, is a free parameter signalling the end of
the resonance region and the beginning of the perturbative QCD domain.
Ideally, predictions should be  stable against reasonable changes in this
parameter. If the hadronic spectral function is known experimentally, 
then the quark mass is extracted with an uncertainty which depends on
the experimental errors of the hadronic data, the value of the continuum
threshold, and the uncertainties in $\Lambda_{QCD}$ and 
the vacuum condensates. If the hadronic spectral function is not known
directly from experiment, it can be reconstructed from information on
resonance masses and widths in the chosen channel. This procedure
introduces two major uncertainties, viz. the value of an overall
normalization, and the question of a potential background (either
constructive or destructive). The choice of a resonance functional
form (e.g. Breit-Wigner), while contributing to the uncertainty, is not
as important. The overall normalization can be constrained considerably
by using chiral-perturbation theory information at threshold. This 
idea was first proposed in \cite{CAD2}, and is now widely adopted in
most applications of QCD sum rules.
However, there still remains the underlying  assumption that chiral symmetry
is realized in the orthodox fashion. If this were not the case, as
advocated e.g. in \cite{STERN}, then the overall normalization would
have to be modified accordingly.
The potential existence of a background, interfering with the resonances,
should be a source of more serious concern. Depending on its size and
sign, it could modify considerably the result for the quark mass, as
will be discussed shortly. 
On the theoretical side, the OPE entails a
factorization of short distance effects (absorbed in the Wilson 
expansion coefficients), and long distance phenomena (represented by the
vacuum condensates). The presence of quark mass singularities of the
form $\ln \; (m_{q}^{2}/Q^{2})$ in the correlator Eq.(6) spoils this 
factorization. Early determinations of $m_{s}$ \cite{NAR} were limited
in precision because of this. A first attempt to deal with this problem
was made in \cite{CAD3}, and a more complete treatement of mass
singularities  is discussed in \cite{CH1}-\cite{J1}.\\ 

In the specific case of the correlator (6), the availability of experimental
data on $K-\pi$ phase shifts \cite{EXP} allows, in principle, for a
reconstruction of the hadronic spectral function, from threshold up to
$s \; \simeq \; \mbox{7 GeV}^{2}$. In both \cite{CH1} and \cite{J1},
the functional form chosen for this reconstruction consisted in a
superposition of two Breit-Wigner resonances, corresponding to the
$K_{0}^{*}$ (1430) and the $K_{0}^{*}$ (1950) \cite{PDG},
normalized at threshold according to 
conventional chiral-symmetry. It was argued in \cite{CH1} and 
\cite{J1} that the non-resonant background implicit in this threshold
normalization was  important to achieve a good fit to the
$K-\pi$ phase shifts. In \cite{CH1} and \cite{J1}, the correlator (6) 
was calculated in perturbative QCD at the 3-loop level, with mass
corrections up to the quartic order, and including non-perturbative
quark and gluon vacuum condensates up to dimension four (the $d=6$
condensates can be safely neglected \cite{J1}). 
Using Laplace transform sum rules, the results
for the strange quark mass thus obtained were
\begin{eqnarray}
m_{s} \; \mbox{(1 GeV)} \; = \; 
\begin{array}{lcl}
171 \pm 15 \; \mbox{MeV} \; \; \; (\cite{CH1})\\[.3cm]
178 \pm 18 \; \mbox{MeV} \; \; \; (\cite{J1})
\end{array}
\end{eqnarray}
The errors reflect uncertainties in the experimental data, in the value
of $\Lambda_{\mbox{QCD}}$ ($\Lambda_{\mbox{QCD}} \simeq$ 200 - 500 MeV),
in the
continuum threshold $s_{0}$ ($s_{0} \simeq \mbox{6 - 7 GeV}^{2}$), and in
the values of the vacuum condensates.\\ 

As mentioned before, an example of a potential systematic error 
affecting these results would be the presence of a background, beyond
the one implicit in the chiral-symmetry normalization of the 
hadronic spectral function at threshold. Obviously, this is not included
in (7). A reanalysis of this QCD sum rule determination  of $m_{s}$
\cite{IT1} has uncovered this uncertainty. In fact, it is claimed                                                        
in \cite{IT1} that by using the Omnes representation to relate the
spectral function to the $K-\pi$ phase shifts, it is necessary to
include a background interacting destructively with the resonances. As
a result, the hadronic spectral function is considerably smaller than
that used in \cite{CH1}-\cite{J1}. This in turn implies smaller values
of $m_{s}$, viz.
\begin{equation}
m_{s} \; \mbox{(1 GeV)} = 140 \pm 20 \mbox{MeV} \; .
\end{equation}

Still another source of systematic uncertainty, this time of a theoretical
nature, has been unveiled in \cite{CH2}. This has to do with the fact
that the QCD expression of the correlator (6) is of the form
\begin{eqnarray}
\psi \; (Q^2) \; \propto \; m_{s}^{2} \; (Q^2) \; \left( 1 +
a_{1} \; (\frac {\alpha_{s}(Q^2)}{\pi}) \; + \right. \nonumber\\
a_{2} \; (\frac {\alpha_{s}(Q^2)}{\pi})^{2} \; + 
a_{3} \; (\frac{\alpha_{s}(Q^2)}{\pi})^{3} \; + \nonumber\\ 
b_{1} \; m_{s}^{2} \; (Q^2) \; (1 + c_{1} \frac{\alpha_{s}(Q^2)}{\pi} \; + 
\cdots) 
+\nonumber\\
\left. b_{2} \; m_{s}^{4}(Q^2) \; (1 + c_{2} \; \frac{\alpha_{s}(Q^2)}{\pi}
\; + \cdots) + \cdots \right)
\end{eqnarray}
Knowing both $m_{s}(Q^2)$  and $\alpha_{s}(Q^2)$ to a given order, say
3-loop, the question is: {\it to expand or not to expand} in 
the inverse logarithms of $Q^{2}$ appearing in (9) ?.
A similar question arises after Laplace transforming 
the correlator, i.e. {\it to expand or not to expand} 
in inverse logarithms of the  Laplace
variable $M_{L}^{2}$ ?. It has been argued in \cite{CH2} that it makes
more sense to make full use of the  perturbative expansions of the quark
mass and coupling (known to 4-loop order), and hence not to expand them
in (9). Numerically, it turns out that the non-expanded expression is
far more stable than the truncated one, when moving from one order in 
perturbation theory to the next. In particular, as shown in \cite{CH2},
logarithmic truncation can lead to sizable overestimates of radiative
corrections. This in turn implies an underestimate of the quark mass.
In fact, after using untruncated expressions, together with the same
hadronic spectral function parametrization as in \cite{CH1}-\cite{J1},
the authors of \cite{CH2} find
\begin{equation}
m_{s} \; \mbox{(1 GeV)} = 203 \pm 20 \mbox{MeV} \; ,
\end{equation}
to be compared with the results (7) obtained from truncated expressions.
Until the question of truncation (or not) becomes satisfactorily settled,
one has to take the value (10) together with (7), and include (8) as well\\.

Finally, I wish to present preliminary results \cite{CAD4} of Laplace
transform QCD sum rules in the strange-pseudoscalar channel, i.e. for
the correlator
\begin{eqnarray}
\psi_{5} (q^2) = i \; \int \; d^4 \; x \; e^{i q x} \nonumber\\
<0|T(\partial^{\mu} \; A_{\mu}(x) \; \partial^{\nu}
\; A_{\nu}^{\dagger}(0))|0> \; ,
\end{eqnarray}
where $A_{\mu}(x) = :\bar{s}(x)  \gamma_{\mu}  \gamma_{5} u(x):$, and
$\partial^{\mu} \; A_{\mu}(x) = m_{s} \; :\bar{s}(x)  i  \gamma_{5}
\;  u(x):$.
The QCD expression of this two-point function is trivially obtained from
that of (6). The hadronic expression, though, is quite different. There
is, at present, preliminary information from tau-decays \cite{ALEPH} in a 
kinematical range restricted by the tau-mass. We have reconstructed the
spectral function, including in addition to the kaon-pole its radial
excitations K(1460) and K(1830), normalized at threshold according to
conventional chiral symmetry. In addition, we have incorporated the
resonant sub-channels $\rho (770)-K$ and $K^{*}(892)-\pi$, which are of
numerical importance (particularly the latter). We find
\begin{equation}
m_{s} \; \mbox{(1 GeV)} = \mbox{165 $ \pm $ 35  MeV} \; ,
\end{equation}
which agrees nicely with results from the scalar channel.
\\
The various QCD sum rule results discussed here can be summarized
in the value
\begin{equation}
m_{s} (1 GeV) = 150 \pm 70 MeV .
\end{equation}
The given error, although rather large, still does not include
the possibility of an unconventional chiral symmetry normalization
of the hadronic spectral functions \cite{STERN}.
Unfortunately, one can not do
better at the moment. The good news are that logarithmic mass
singularities can be dealt with satisfactorily \cite{CH1}-\cite{J1},
and that the 4-loop ($\cal O (\alpha_{s}^{3})$) perturbative
corrections are small. This reduces considerably the 
theoretical uncertainties of the method. However, some still 
remain, largely in the issue of the expansion (or not)
in inverse logarithms, when taking products of (running) quark
masses and the coupling constant. They affect $m_{s}(1 GeV)$ by
up to 20 - 25 \%. The major source of error, though, lies in the
hadronic sector. Direct and accurate measurements of the relevant
spectral functions appear as the only hope to reduce the uncertainty
in the quark masses.\\

Acknowledgements: The author wishes to thank Harald Fritzsch for
his kind hospitality in M\"{u}nchen , and the organizers of WIN97
for a great workshop.\\

\end{document}